\documentclass[preprint,showpacs,preprintnumbers,amsmath,amssymb]{revtex4}

\usepackage{graphicx}
\usepackage{dcolumn}
\usepackage{bm}

\begin{document}


\title{Chaotic behavior in the accretion disk}

\author{LIU Lei}
 \email{liulei@mail.iap.ac.cn}
 \altaffiliation[Also at ]{Graduate University of the Chinese Academy of Sciences,
  Beijing 100049, China}
\author{HU Fei}
\affiliation{Institute of Atmospheric Physics, Chinese Academy of
Sciences, Beijing 100029, China}

\date{\today}

\begin{abstract}
The eccentric luminosity variation of quasars is still a mystery.
Analytic results of this behavior ranged from multi-periodic
behavior to a purely random process. Recently, we have used
nonlinear time-series analysis to analyze the light curve of 3C 273
and found its eccentric behavior may be chaos [L. Liu, Chin.\ J.\
Astron.\ Astrophys. \textbf{6}, 663 (2006)]. This result induces us
to look for some nonlinear mechanism to explain the eccentric
luminosity variation. In this paper, we propose a simple non-linear
accretion disk model and find it shows a kind of chaotic behavior
under some circumstances. Then we compute the outburst energy
$\triangle F$, defined as the difference of the maximum luminosity
and the minimum luminosity, and the mean luminosity $\langle F
\rangle$. We find that $\triangle F\sim \langle F \rangle^{\alpha}$
in the chaotic domain, where $\alpha\approx 1$. In this domain, we
also find that $\langle F \rangle \sim M^{0.5}$, where $M$ is the
mass of central black hole. These results are confirmed by or
compatible with some results from the observational data analysis
[A. J. Pica and A. G. Smith, Astrophys.\ J. \textbf{272}, 11 (1983);
M. Wold, M. S. Brotherton and Z. Shang, Mon.\ Not.\ R.\ Astron.\
Soc. \textbf{375}, 989 (2007)].
\end{abstract}

\maketitle

\section{Introduction}

Since the discovery of quasar in 1963 \cite{sh63}, the luminosity
variation has played an important role in our understanding of its
nature. Although it has been subjected to extensive analysis, there
is no generally accepted method of extracting the information in the
light curve. Results of analysis of the light curve ranged from
multi-periodic behavior \cite{kunkel,jurkev,shk88,lin01} to a purely
random process \cite{ms68,to70,fu75}. On the basis of these
analysis, many theories which ranged from periodic mechanisms
\cite{lin01,ar99,rcaf00} to superimposed random events such as the
so-called Christmas tree model \cite{moore}, have been proposed.
Then whatever does this seemingly random light curve tell us? Could
this seeming randomness be some behavior other than multi-periodic
or purely random?

Also in 1963, Edward Lorenz published his monumental work entitled
{\it Deterministic Nonperiodic Flow} \cite{lorenz}. In this paper,
he found a strange behavior which can appear in a deterministic
non-linear dissipative system, which seems random and unpredictable,
and is called Chaos. Chaotic behavior is not multi-periodic because
it has a continuous spectrum. Useful information can not always be
extracted from the power spectrum of chaotic signal. On the other
hand chaotic behavior is not random either because it can appear in
a completely deterministic system. The concept of attractor is often
used when describing chaotic behaviors. As the dissipative system
evolves in time, the trajectory in state space may head for some
final region called attractor. The attractor may be an ordinary
Euclidean object or a fractal \cite{feder} which has a non-integer
dimension and often appears in the state space of a chaotic system.
For many practical systems, we may not know in advance the required
degrees of freedom and hence can not measure all the dynamic
variables. How can we discern the nature of the attractor from the
available experimental data? Packard et al. \cite{pcf80} introduced
a technique which can be used to reconstruct state-space attractor
from the time series data of a single dynamical variable. Moreover,
the correlation integral algorithm subsequently introduced by
Grassberger and Procaccia \cite{gp83} can be used to determine the
dimension of the attractor embedded in the new state space. These
techniques constitute a useful diagnostic method of chaos in
practical systems.

Therefore, the above-mentioned diagnostic methods have been used to
analyze the light curve of 3C 273, and it has been found that the
eccentric luminosity variation of 3C 273 may be a kind of chaotic
behavior \cite{liu06}. This result tells us that non-linear may play
a very important role in the nature of quasar. Then whatever role
does the non-linear play? Could the non-linear mechanisms help us
understand the eccentric behavior in the luminosity variation of
quasars?

For the sake of keys, we in this paper propose a simple non-linear
accretion disk (NAD) model, by borrowing basic ideas from the
standard accretion disk (SAD) \cite{ss73,pringle}. We then find that
with some parameters the disk shows a kind of chaotic behavior.
Although our model is based on the SAD, there are two main
differences between them. First, SAD is a stable disk which can not
be used to explain the eccentric luminosity variation. Second, SAD
uses the complex differential equations of fluid dynamics to
describe its evolution and NAD uses iterated equations. The latter
simplifies the problem to a great degree. In the following sections,
you will see very simple non-linear terms can produce extremely
complex light curves. What's more, NAD is more natural than many
other models on luminosity variation because our world is a
non-linear world and the NAD aims to explore the non-linear effects
on the quasar. Overall, our results provide a new view of the nature
of the eccentric luminosity variation and may be helpful in the
future study of quasars.

\section{Description of the Chaotic Accretion Disk Model}
There are always two kinds of factors which can be classified by
their contrary contributions to the behavior of the non-linear
chaotic systems. For example, in the famous Logistic model
\cite{may}, one factor is related with the plenty food and/or space
etc., and the other is related with the overpopulation and disease.
The formal can make the population grow, and the latter can make the
population decrease. The interaction of the two factors can lead to
the chaotic variation of the population under some circumstances.
The other example is the Lorenz model \cite{lorenz}. In this model,
the two kinds of contrary factors are the heating of the earth, and
the viscous stresses and the gravity. The heating of the earth can
make the air rise. However the viscous stresses and the gravity
always prevent the air rising. The interaction between them will
lead to the chaotic behavior of the air.

We note that the standard accretion disk (SAD) also has two kinds of
contrary factors as in the above models. On the one hand, the
viscous stresses will make the gas rotate slowly around the black
hole. On the other hand, the gas in the disk will flow towards black
hole, for its centrifugal force and gravity is not in equilibrium
due to viscous stresses. The gravitational potential energy will
convert to the kinetic energy on the process of flowing towards the
black hole which makes the gas rotate fast. Thus the viscous
stresses and the gravity are the two contrary factors in the SAD.
This is the start point of our non-linear accretion disk (NAD)
model. By borrowing these basic ideas, we now construct the NAD
model as follows.

\begin{figure*}
  \includegraphics*{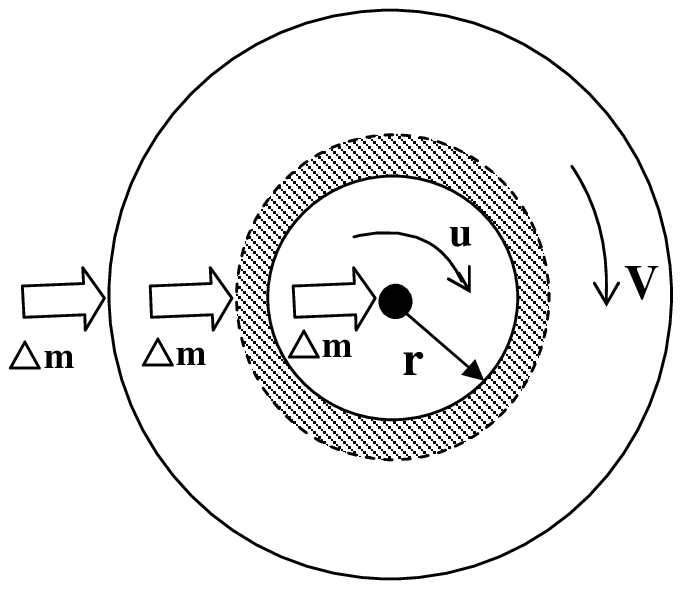}
  \caption{\label{fig:nad} The NAD model. The black dot at the centre of this figure
  represents a black hole and the shadow region is the boundary layer.
  Broad arrays represent the flow of mass towards black
  hole.}
\end{figure*}

We consider an accretion disk which has only two layers of gas, as
in Fig.~\ref{fig:nad}. It is supposed that (a) the thickness of each
layer is very small that the tangential velocity in each layer is
slightly different from its average tangential velocity. That is to
say, we only need one velocity to describe the circular motion of
one layer; (b) the height of disk $H$ is very small that the motion
is almost two-dimensional; (c) the average tangential velocity of
outer layer is a constant and equals its average Kepler's velocity
$V$. Due to the viscous stresses, the average tangential velocity of
inner layer $u$ does not equal its average Kepler's velocity $U$,
and is a time variable; (d) the inner layer interacts with outer
layer by viscous stresses, but it does not interact with black hole;
(e) during small interval $\triangle \tau$, the mass $\triangle m$
flows across any circle (all the centers of circles are on the
center of black hole) is the same. If $U^2> u^2$, the centrifugal
force is less than the gravity and the gas will flow towards black
hole. Define $\triangle m>0$ in such a case. If $U^2< u^2$, the gas
will flow far from the black hole and $\triangle m <0$. When $U^2 =
u^2$, $\triangle m=0$. Thus, we simply suppose that $\triangle m
\propto £¨U^2-u^2£©$; (f) there is a boundary layer between outer
layer and inner layer. The thickness of this boundary layer $h$ is
very small comparing with the radius $r$ of the inner layer's outer
edge. When gas in the outer layer flows across the boundary layer,
the average tangential velocity will change from $V$ to $U$; (g) Any
relativistic effect is not considered here. All above are the basic
assumptions of our NAD model.By these assumptions, the equations of
motion are derived as follows.

At the $\mathrm{n^{th}}$ moment, the viscous stresses acting on the
inner layer is,
\[
f_{\mathrm{n}}=2 \pi rH\mu\frac{u_{\mathrm{n}}-V}{h},
\]
where $u_{\mathrm{n}}$ is the average tangential velocity of inner
layer at the $\mathrm{n^{th}}$ moment and $\mu$ is the viscosity.
Then we have,
\[
2 \pi
rH\mu\frac{u_{\mathrm{n}}-V}{h}=m\frac{u_{\mathrm{n}}-{u_{\mathrm{n+1}}}^-}{\triangle
\tau},
\]
where $m$ is the mass of inner layer, ${u_{\mathrm{n+1}}}^-$ is the
average tangential velocity of inner layer at $\mathrm{(n+1)^{th}}$
moment due to the action of viscous stresses, and $\triangle \tau$
is the time interval between $\mathrm{n^{th}}$ and
$\mathrm{(n+1)^{th}}$ moments. Thus it can be obtained that,
\begin{equation}
{u_\mathrm{n+1}}^-=u_\mathrm{n}-\frac{2\pi rH\mu\triangle
\tau}{hm}(u_\mathrm{n}-V) \label{eq:one}.
\end{equation}

When ${u_\mathrm{n}}^2 < U^2$, the mass $\triangle m_\mathrm{n}$ in
inner layer will flow into black hole, which causes the same mass to
flow across boundary layer into inner layer. When the gas flows
across boundary layer, its velocity will change from $V$ to $U$.
According to the assumption (e),
\begin{equation}
\frac{\triangle
m_\mathrm{n}}{m}=C\Big(1-\frac{{u_\mathrm{n}}^2}{U^2}\Big)
\label{eq:two}£¬
\end{equation}
where $C$ is a dimensionless parameter. At the $\mathrm{(n+1)^{th}}$
moment, the average tangential velocity due to the flow of mass
towards black hole is
\[
{u_\mathrm{n+1}}^+=\frac{(m-\triangle
m_\mathrm{n})u_\mathrm{n}+\triangle
m_\mathrm{n}U}{m}\\
=u_\mathrm{n}+(U-u_\mathrm{n})\frac{\triangle m_\mathrm{n}}{m}.
\]
According to Eq.~(\ref{eq:two}), we have
\begin{equation}
{u_\mathrm{n+1}}^+ =
u_\mathrm{n}+C\Big(1-\frac{{u_\mathrm{n}}^2}{U^2}\Big)(U-u_\mathrm{n})
\label{eq:third}.
\end{equation}

When ${u_\mathrm{n}}^2 > U^2$, there is only mass flowing out of
inner layer, which will not change the average tangential velocity
of inner layer. Thus,
\begin{equation}
{u_\mathrm{n+1}}^+ = u_\mathrm{n} \label{eq:four}.
\end{equation}

Finally, the average tangential velocity of inner layer at the
$\mathrm{(n+1)^{th}}$ moment is
\[
u_\mathrm{n+1}=\\{u_\mathrm{n+1}}^+ + {u_\mathrm{n+1}}^-
-u_\mathrm{n}.
\]
According to Eqs.~(\ref{eq:one}), (\ref{eq:third}) and
(\ref{eq:four}), we have
\[
u_\mathrm{n+1}=\left\{\begin{array}{ll}
u_\mathrm{n}+C(1-{u_\mathrm{n}}^2/U^2)(U-u_\mathrm{n})-(2\pi
rH\mu\triangle \tau/hm)(u_\mathrm{n}-V)& \textrm{when
${u_\mathrm{n}}^2\leq
U^2$}\\
u_\mathrm{n}-(2\pi rH\mu\triangle \tau/hm)(u_\mathrm{n}-V)& \textrm
{when ${u_\mathrm{n}}^2>U^2$} \end{array}\right.
\]
Normalizing above equations, we have
\begin{equation}
\overline{u_\mathrm{n+1}}=\left\{\begin{array}{ll}
\overline{u_\mathrm{n}}+C(1-\overline{u_\mathrm{n}}^2)(1-\overline{u_\mathrm{n}})
-A(\overline{u_\mathrm{n}}~-B)&
\textrm{when $\overline{u_\mathrm{n}}^2\leq
1$}\\
\overline u_\mathrm{n}-A(\overline u_\mathrm{n}-B)& \textrm{when
$\overline{u_\mathrm{n}}^2>1$}\end{array} \right.
\label{eq:seven}
\end{equation}
where $\overline{u_\mathrm{n}}=u_\mathrm{n}/U$ and dimensionless
parameters
\begin{equation}
A=\frac{2\pi rH\mu \triangle\tau}{hm}, \hspace{0.4cm} B=\frac{V}{U}
\label{eq:five}.
\end{equation}

We assume that the whole heat produced by viscous stresses was
radiated out along the direction which is perpendicular to the
accretion disk. Thus the luminosity at the $\mathrm{n^{th}}$ moment
is
\[
F_\mathrm{n}=f_\mathrm{n}\cdot(u_\mathrm{n}-V)=2\pi
rH\mu\frac{(u_\mathrm{n}-V)^2}{h}.
\]
Normalizing above equation, we have
\begin{equation}
\overline{F_\mathrm{n}}=A(\overline{u_\mathrm{n}}-B)^{2}
\label{eq:eight},
\end{equation}
where the dimensionless luminosity is
\[\overline{F_\mathrm{n}}=\frac{F_\mathrm{n}\triangle \tau}{mU^2}.\]

From Eq.~(\ref{eq:five}), we can see that $A$ is related to the
viscous and geometry properties of the NAD. For avoiding the
singularities in NAD model, the inequality
\[
\left|\frac{d \overline u_\mathrm{n+1}}{d \overline
u_\mathrm{n}}\right|<1
\]
must be satisfied when $\overline {u_\mathrm{n}}^2>1$. Then we can
obtain that
\[
0<A<2.
\]
Under the assumption (a) and Eq.~(\ref{eq:five}), we have
\[
B=\sqrt\frac{3r_g}{r}=\sqrt\frac{6GM}{c^2r},
\]
where $r_g$ is the Schwarzschild radius, $c$ is the velocity of
light and $M$ is the mass of the black hole. Clearly, $B$ is related
to the mass of the black hole and the geometry properties of the NAD
and
\[
0<B<1.
\]
The parameter $C$ can also be related to some physical quantities.
The distance that the mass $\triangle m$ at the outer edge of inner
layer travels towards black hole during $\triangle \tau$ is,
\[
\triangle s=\frac{U^2-u^2}{2r}\triangle \tau^2.
\]
Then we can be obtained that
\begin{equation}
\frac{\triangle m}{m}=\frac{\pi \rho H U^2 \triangle
\tau^2}{m}(1-\overline u^2) \label{eq:six}.
\end{equation}
Comparing Eqs.~(\ref{eq:six}) with (\ref{eq:two}), we find that
\[
C=\frac{\pi \rho H U^2 \triangle \tau^2}{m}=\frac{U^2 \triangle
\tau^2}{r^2}=\frac{c^2 \triangle \tau^2}{6r^2}.
\]
We can see that $C$ is only related to the geometry properties of
the NAD. Here we assume that the mass of inflow or outflow during
$\triangle t$ do not exceed the mass of the inner layer, that is,
\begin{equation}
-1<\frac{\triangle m_\mathrm{n}}{m}<1 \label{eq:nine}.
\end{equation}
For the inequality $\triangle m_\mathrm{n}/m<1$ is always satisfied
when $\overline{u_\mathrm{n}}^2\leq 1$, we choose
\[
0<C<1
\]
in the following discussion. From above discussions, we can see that
only $A$ is related to the viscous property of NAD, and only $B$ is
related to the mass of black hole. Thus by tuning corresponding
parameter, we can easily know the effect of the variation of some
physical property on the behavior of the NAD and its light curve.
Some results are given in the following sections.

\section{Results}

\begin{figure*}
 \resizebox{7.5cm}{!}{\includegraphics{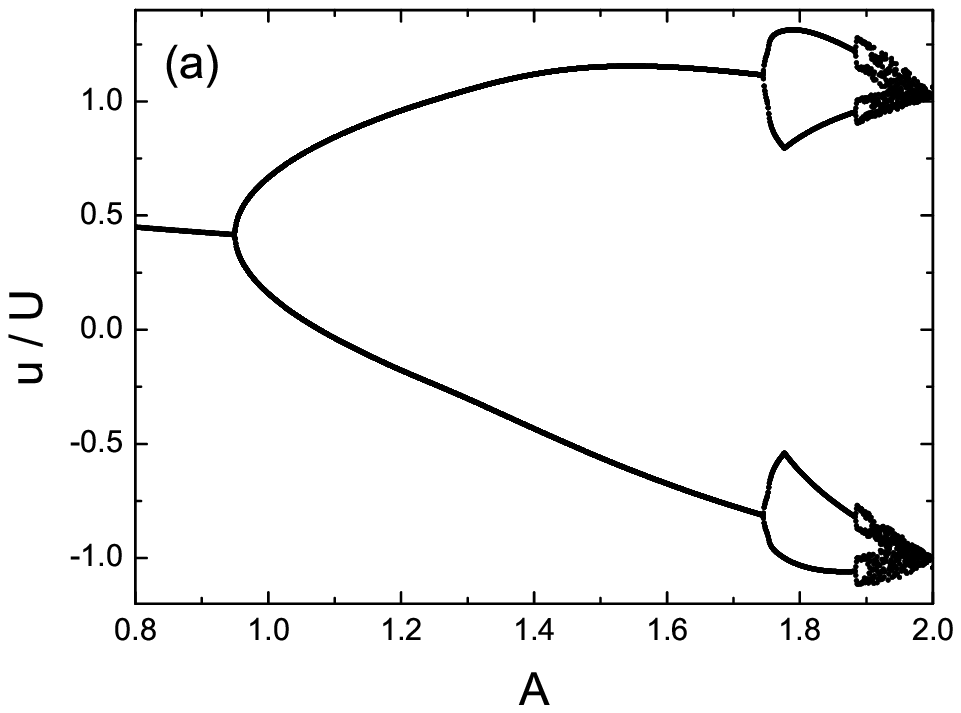}}
  \hspace{0.5cm} \resizebox{7.5cm} {!}{\includegraphics{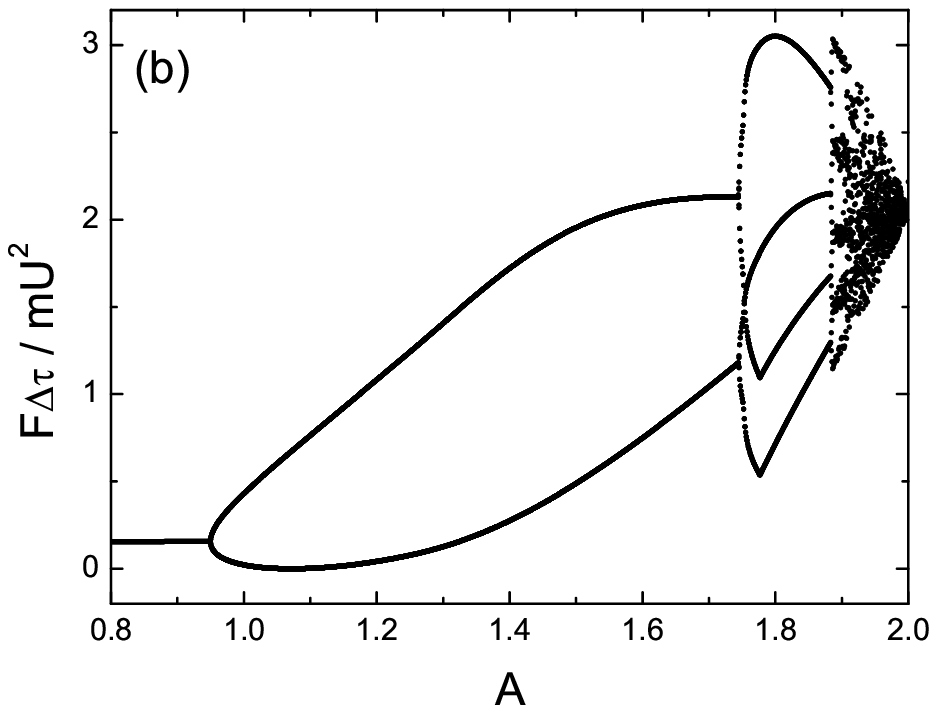}}
    \resizebox{7.5cm} {!}{\includegraphics{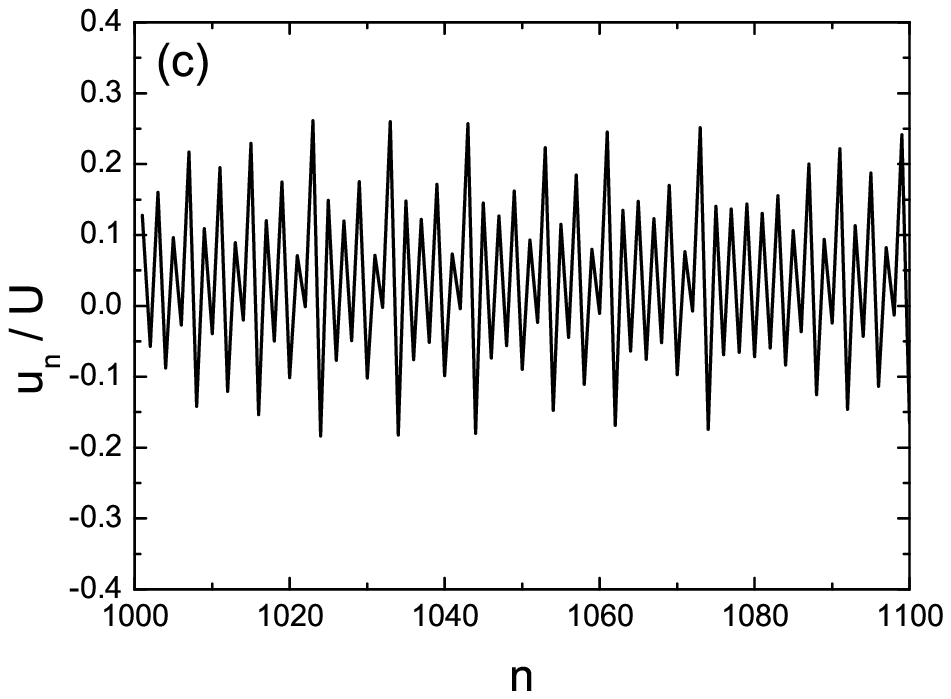}}
  \hspace{0.5cm} \resizebox{7.5cm} {!}{\includegraphics{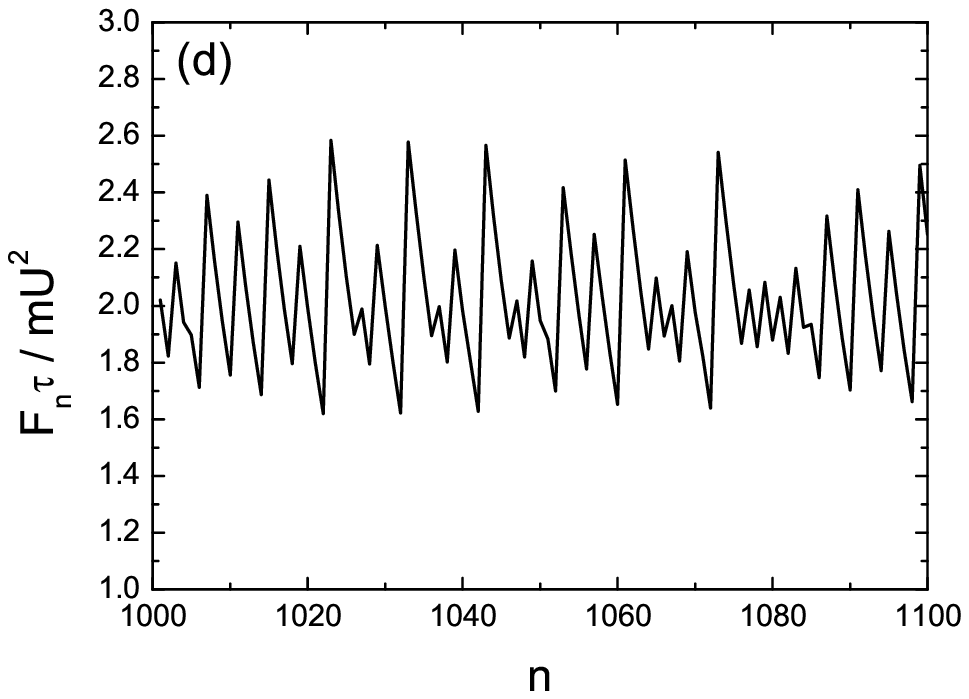}}
  \caption{\label{fig:one} (a) and (b) are bifurcation diagrams of NAD model
  when $B=0.01$ and $C=0.8$. (a) is computed by picking a value of $A$ and a
  starting point of $u_0=1$, iterating Eq.~(\ref{eq:seven}) 1000 times
  to allow the trajectory to approach the attractor and plotting
  the next 5000 values of $u$. Then (b) is computed by Eq.~(\ref{eq:eight}).
  (c) is the average tangential
  velocity $u_\mathrm{n}$ of inner disk as a function of time
  and is computed by picking $A=1.95$, $B=0.01$ and $C=0.8$, iterating
  Eq.~(\ref{eq:seven})
  1000 times from initial point $u_0=1$ and then plotting the next 100 values of $u$.
  For clearly showing its chaotic behavior,
  all the values of $u_\mathrm{n}$ are minus 0.9 if they are greater than zero;
  conversely, they are plus 0.9. (d) is the light curve
  computed by Eq.~(\ref{eq:eight}) also with $A=1.95$, $B=0.01$ and
  $C=0.8$.}
\end{figure*}
\begin{figure*}
\resizebox{7.5cm}{!}{\includegraphics{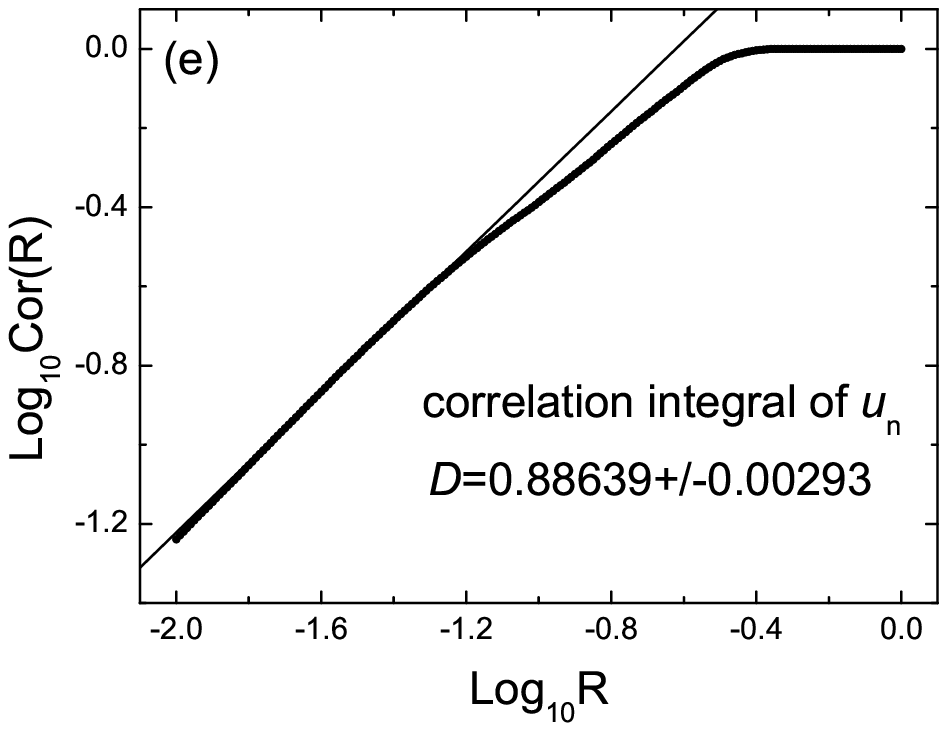}}
\hspace{0.5cm} \resizebox{7.5cm}{!}{\includegraphics{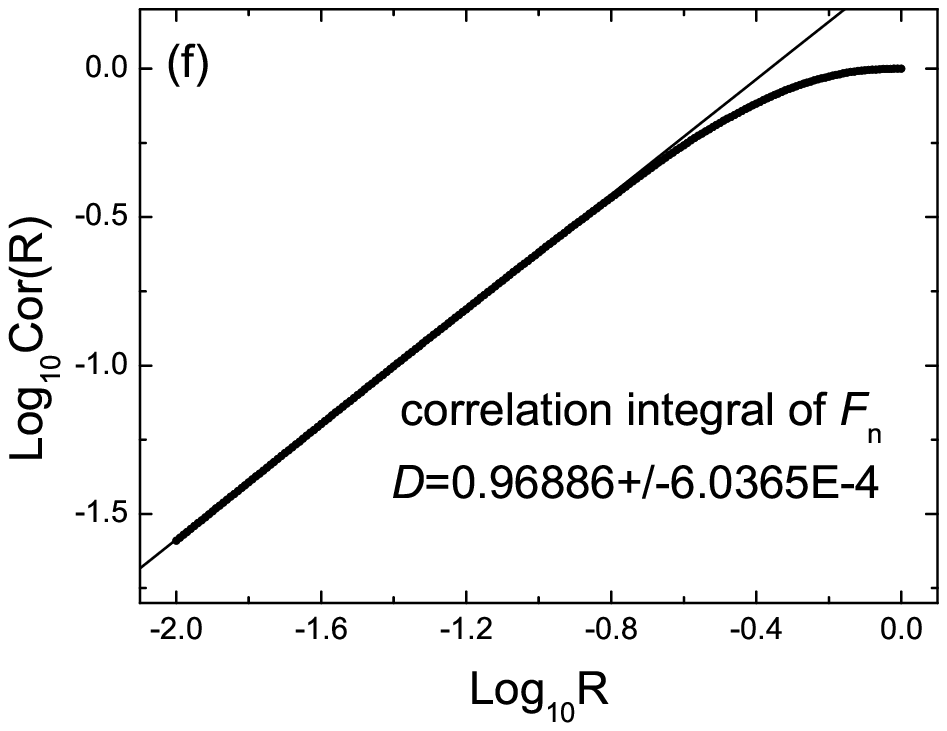}}
\caption{\label{fig:two} (e) and (f) are correlation integrals when
$A=1.95$, $B=0.01$ and $C=0.8$. They are computed by using 5000 data
after 1000 iterations as in Fig.~\ref{fig:one}(a)--(d), and the
corresponding correlation dimension is given in each plot.}
\end{figure*}

In this section, some properties of our model will be discussed by
numerical computation. First,we will discuss the long-term chaotic
behavior of NAD with different viscous properties. As discussed in
the above section, the viscous properties of NAD are just related to
the parameters $A$. Thus in our following discussion $B$ and $C$ are
fixed and $A$ is an adjustable parameter. Fig.~\ref{fig:one}a and
Fig.~\ref{fig:one}b are typical bifurcation diagrams when $B=0.01$
and $C=0.8$. They are very similar to the Logistic bifurcation
diagram \cite{hil94}. From these figures, we can see that the
behavior of NAD will become more and more complex with the increase
of viscosity. Specially, there is a piece of vague area on the right
side of each figure. The behavior of $u$ in Fig.~\ref{fig:one}a (or
$F$ in Fig.~\ref{fig:one}b) in this area will be chaotic.
Fig.~\ref{fig:one}c is the evolution of the average tangential
velocity $u$ of the inner layer and Fig.~2d is the light curve
generated by NAD model. Both of them are plotted when
$A=1.95$,$B=0.01$,$C=0.8$, with which the system shows a kind of
chaotic behavior. Here we must note that the choice of the
parameters is not arbitrary. Any choice of the parameters should
guarantee that Eq.~(\ref{eq:nine}) must be satisfied. By many
numerical computations, we find that when $A$ is greater than about
$1.9$ and $C$ is greater than about $0.6$, Eq.~(\ref{eq:nine}) is
satisfied and the luminosity is chaotic.

The correlation integral \cite{gp83} is also used to analyze our
model. Let $X_{1},X_{2},\ldots,X_{N}$ be samples of a physical
variable ($u$ or $F$ in our model) at the $\mathrm{i^{th}}$ moment,
$\mathrm{i}=1,2,\ldots,\mathrm{N}$. The correlation integral is
defined as
\[
\mathrm{Cor(R)}=\frac{1}{\mathrm{N(N-1)}}\sum_{\mathrm{i}=1}^{\mathrm{N}}\sum_{\mathrm{j}=1,\mathrm{j}\neq
\mathrm{i}}^{\mathrm{N}}\theta(\mathrm{R}-|X_{\mathrm{i}}-X_{\mathrm{j}}|),
\]
where $\theta(x)$ is the Heaviside function,
\[
\theta(x)=\left \{
\begin{array}{ll}
1 & x \geq 0\\
0 & x<0.
\end{array} \right.
\]
In Ref.~\onlinecite{gp83}, they have studied many chaos models and
found that
\[
\mathrm{Cor(R)}\propto \mathrm{R}^{D},
\]
where $D$ is called correlation dimension. Strictly $D$ is not the
dimension of the attractor, but is very close to it. According to
chaos and non-linear dynamics, if an attractor for a dissipative
system has a non-integer dimension, then the attractor is a chaotic
attractor \cite{hil94}. Therefore, the correlation integral is a
useful diagnostic tool of chaos. However, itself does not have any
physical meaning. Fig.~\ref{fig:two}e and Fig.~\ref{fig:two}f are
typical results of correlation integral when $A=1.95$, $B=0.01$,
$C=0.8$. From these results, we can see that attractors in the state
space of $u$ and $F$ are indeed chaotic attractors.

\begin{figure*}
\resizebox{10cm}{!}{\includegraphics{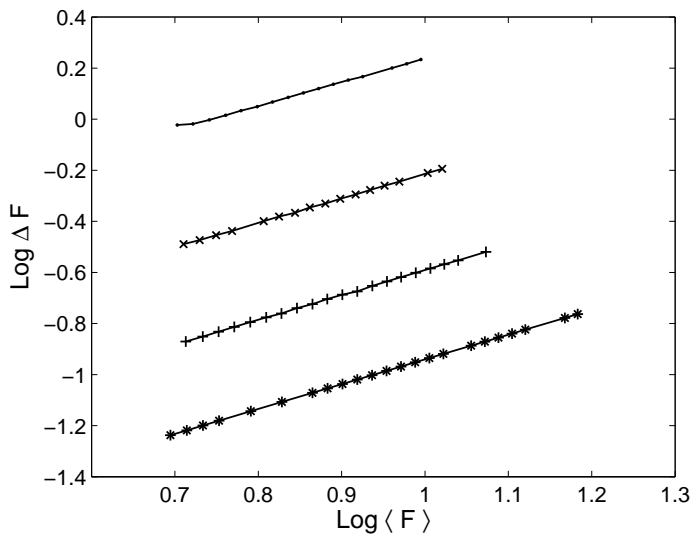}}
\caption{\label{fig:outburst} The outburst energy $\triangle F$
 as a function of the mean luminosity $F_m$. Here, we choose four groups of
 parameters. In each group, parameters $A$ and $C$ are fixed,
 whereas $B$ varies in the interval $(0,1)$, so long as the light curve is chaotic
 and Eq.~(\ref{eq:nine}) is satisfied. The samples of the light curve
 are produced as in Fig.~\ref{fig:one} and the results are plot with lines and different
 symbols. The corresponding parameters and the slope of the line $\alpha$ are given
 as follows. Point, $A=1.95,C=0.8,\alpha \approx 0.93$; cross, $A=1.97,C=0.8,\alpha \approx 0.96$;
 plus, $A=1.98,C=0.8,\alpha \approx 0.97$; star, $A=1.98,C=0.6,\alpha \approx
 0.97$.}
\end{figure*}
\begin{figure*}
\resizebox{10cm}{!}{\includegraphics{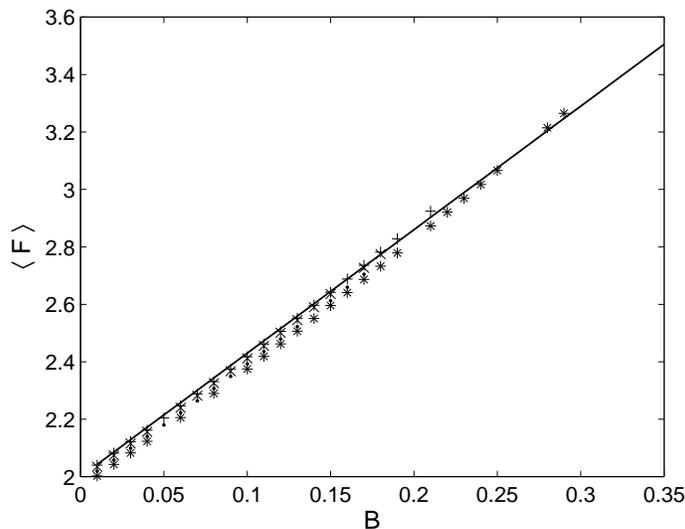}}
\caption{\label{fig:mass} The mean luminosity as a function of the
parameter B. The parameters, the samples of the light curve and the
corresponding symbols are the same as Fig.~\ref{fig:outburst}. The
line in the plot is $\langle F \rangle= 4.3B+2$.}
\end{figure*}

We then discuss the relationship of the outburst energy and the mean
luminosity. The outburst energy $\triangle F$ is defined as the
difference between maximum luminosity $\overline F(\textrm{max})$
and minimum luminosity $\overline F(\textrm{min})$, that is,
\[
\triangle F=\overline F(\textrm{max})-\overline F(\textrm{min}).
\]
This definition is the same as in Ref.~\onlinecite{ps83}. Here, we
fix the parameters $A$ and $C$ and compute the luminosity with
different $B$. For a guarantee of satisfying Eq.~(\ref{eq:nine}) and
the chaotic behavior in the light curve, we choose $A>1.9$ and
$C>0.6$, as was stated above. Then we compute the outburst energy
and and the mean luminosity $\langle F \rangle$, and find that
\begin{equation}
\triangle F\sim  \langle F \rangle^{\alpha} \label{eq:ten},
\end{equation}
where $\alpha \approx 1$, as in Fig.~\ref{fig:outburst}. This result
is very similar with the observational facts found in the Fig.~7 of
Ref.~\onlinecite{ps83}, where they found that for most sources it
appears that the outburst energy scales with the mean luminosity.
Additionally, we should note that $B$ is related to the mass of the
black hole. Thus this find suggests that the mass of black hole may
not affect the nature of the properties of the luminosity variation.
Quasars with big black holes will obey the same rules of the
luminosity variation as the small ones. By fixing $A$ and $C$, the
relationship of mean luminosity and the mass of black hole is also
discussed. We find that the mean luminosity scales with $B$, as in
Fig.~\ref{fig:mass}. With Eq.~(\ref{eq:five}), we can conclude that,
\begin{equation}
\langle F \rangle\sim M^{1/2} \label{eq:eleven}.
\end{equation}
Combining Eq.(\ref{eq:ten}) with Eq.~(\ref{eq:eleven}), we
immediately obtain that,
\begin{equation}
\triangle F \sim M^{\beta}\label{eq:twelve},
\end{equation}
where $\beta\approx 1/2$. Ref.~\onlinecite{wbs07} have reported that
it is evident that the sources displaying largest variability
amplitudes have, on average, higher black hole masses, although
there is no linear relationship between them. Eq.~(\ref{eq:twelve})
is compatible with their results.

\section{Discussion and Conclusion}

In this paper we propose a non-linear accretion disk model (NAD)
which can be used to describe the chaotic behavior observed in the
light curve of quasar 3C 273 \cite{liu06}. We note that the
tangential velocity of accretion disk is influenced by two factors.
One is the viscous stresses and the other is the flow of mass
towards black hole. Under some circumstances, one of them may reduce
the velocity and the other may increase it. Because of non-linear,
the winner of gaining upper hand among them always vary with time.
Then a kind of irregular oscillation of accretion disk would come
out. It finally leads to the chaotic luminosity variation.

Although any complex multi-periodic mechanism and unnatural random
event is not included, very simple non-linear terms in our model can
also produce extremely complex behavior of the accretion disk.
Meanwhile, for understanding the underlying non-linear nature of the
chaotic behavior of the light curve, we ignore many physical
details, including some may be very important to the behavior of the
quasars. For example, the effects of the thermodynamics \cite{ny94},
the interaction between the inner edge of the inner layer and the
black hole \cite{kh02}, and the details of the dynamics happening in
the boundary layer are not considered here. Additionally, we use the
iterated equations instead of the complex equations of fluid
dynamics. All above-mentioned facts simplify the problem greatly.
However, the simplification does not weaken the validity of our
model on explaining the eccentric luminosity variation. The model
can shows two rules in the chaotic domain. One is the outburst
energy $\triangle F\sim \langle F \rangle^{\alpha}$, where $\langle
F \rangle$ is the mean luminosity and $\alpha \approx 1$; the other
is $\langle F \rangle\sim M^{0.5}$, where $M$ is the mass of central
black hole. There rules are confirmed by or compatible with the
observational data analysis \cite{ps83,wbs07}.

The standard accretion disk model (SAD) is always be recognized as a
standard picture of the AGNs, for it can produce extreme energy that
be observed in the AGNs \cite{bell}. Our NAD model just borrows the
basic idea of the SAD. Thus, we conclude that the chaotic behavior
presenting in NAD model can be seen in many eccentric light curves
of the AGNs. Some authors have reported that other AGNs may have
chaotic behavior in their eccentric light curves \cite{mhak06}.
However, we hope that there will be more evidences to support or
oppose the assertions of our model, especially the quantitative
rules predicted in this paper.

Overall, our results reveal that the non-linear plays a key role in
the cause of the eccentric luminosity variation of quasars. The
omnipresent non-linear is very important in our understanding of
this complex world, as chaos and non-linear dynamics tells us. Thus
it is nature to study the luminosity variation of quasars from the
view of non-linear. And we hope chaos and non-linear dynamics may be
helpful in the further study of quasars.

\begin{acknowledgments}
LIU Lei thanks Professor Meng Ta-Chung for his kind help to
understand the chaos and non-linear dynamics. Many thanks go to Dr.
CHENG Xue-Ling, GUO Long, ZHU Li-Lin and DING Heng-Tong for helpful
discussions and ZHU Yan for improving the plot. This research is
supported by the National Nature Science Foundation of China under
Grant No. 40405004.
\end{acknowledgments}

\newpage 
\bibliography{myfile}

\end{document}